\documentclass[aps,preprint, double-spaced,
showpacs,superscriptaddress,floatfix,nofootinbib]{revtex4}%
\usepackage{amssymb}
\usepackage{graphicx}
\usepackage[english]{babel}%
\usepackage{amsmath}%
\usepackage{amsfonts}
\newcommand{\be}{\begin{equation}}
\newcommand{\ee}{\end{equation}}

\begin{document}
\title{Geometric Scaling in the Quantum Hall System}
\author{C.A. L\"utken}
\affiliation{Theory Group, Department of Physics, University of Oslo}
\author{G.G. Ross}
\affiliation{Rudolf Peierls Centre for Theoretical Physics, Department of Physics,
University of Oxford}

\begin{abstract}
The transitions between neighbouring plateaux in the quantum Hall system are
observed to follow \textquotedblleft anti-holomophic\textquotedblright%
\ scaling with \textquotedblleft superuniversal\textquotedblright\ scaling
exponents, showing that the system contains an emergent sub-modular discrete
symmetry and a holomorphic structure at low energies. We identify a class of
effective scaling models consistent with this data, which is parametrized by
the complex structure of a torus with a special spin structure, in which only
the number of fermions $(c)$ remains undetermined. For $c=2$ this gives the
superuniversal anti-holomorphic scaling potential previously inferred from
data, with scaling exponent $\nu\approx2.6$, in reasonable agreement with
available scaling data.

\end{abstract}

\pacs{73.20.-r}
\maketitle

It has previously been pointed out that the fixed point structure and unusual
scaling behaviour observed in the quantum Hall system suggests a unified
approach to the integer and fractional quantum Hall effects \cite{LR}. Scaling
is observed in the transition between neighbouring plateaux, and the scaling
exponents appear to be independent of which transition is examined
(``superuniversality''). The conductivities $\sigma^{xx}$ and $\sigma^{xy}$
are the effective coupling constants in the low-energy effective field theory
(EFT) of quantum charge transport in an external magnetic field in a
disordered medium. We can therefore treat these scaling phenomena as critical
points of the renormalization group (RG) beta-functions $\beta^{xx} =
d\sigma^{xx}/dt$ and $\beta^{xy} = d\sigma^{xy}/dt$, where $t$ is the dominant
scale parameter \cite{K, LLP, LR}.

Superuniversality is automatic if the EFT posesses a kind of \textquotedblleft
complexified Kramers-Wannier duality\textquotedblright, identified in
\cite{LR} as the symmetry group $\Gamma_{0}(2)$, which is generated by
translations $T:\sigma=\sigma^{xy}+i\sigma^{xx}\rightarrow\sigma+1$ and the
\textquotedblleft complexified duality transformation for
fermions\textquotedblright\ $ST^{2}S:\sigma\rightarrow\sigma/(1-2\sigma)$. The
complexified beta-function $\beta^{\sigma}=d\sigma/dt=\beta^{xy}+i\beta^{xx}$
transforms as a contravariant vector under $\Gamma_{0}(2)$-transformations
\cite{BL}, but this is not sufficient information usefully to constrain the
scaling theory. Recently it was pointed out that the scaling data provide
additional information \cite{LR2}. When the scaling exponents for $\sigma
^{xy}$ and $\sigma^{xx}$ are defined geometrically they appear to be of equal
magnitude, which strongly suggests that the effective scaling theory contains
a holomorphic structure. Taken together these constraints are so strong that
the beta-function in the scaling region was found to be unique up to
normalization: $\beta_{\sigma}\varpropto\partial_{\sigma}\varphi(\sigma)$,
where
\begin{equation}
\varphi(\sigma)=\ln\Delta(2\sigma)-\ln\Delta(\sigma),\label{eq:data}%
\end{equation}
and $\Delta(\sigma)$ is the modular cusp form of weight $12$. This scaling
form is a complex analytic (holomorphic), everywhere non-singular and
non-vanishing function scalar (weightless) under $\Gamma_{0}(2)$. It plays the
role of Zamolodchikov's RG-potential (C-function) in his celebrated C-theorem
\cite{Zam}, which spells out what needs to be done in simple geometrical
terms: we need a model containing sufficient dynamical information to yield
the covariant gradient flow \textit{physically normalized}, $\beta
_{\overline{\sigma}}=\partial_{\overline{\sigma}}C$, which together with the
physical parameter space metric $G_{\sigma\overline{\sigma}}$ gives the
physical (contravariant) beta-function: $\beta^{phys}=\beta^{\sigma}%
=G^{\sigma\overline{\sigma}}\beta_{\overline{\sigma}}/12$.

While a mathematical analysis of eq.~(\ref{eq:data}) is beyond the scope of
this letter, we pause here briefly to remark on the simplicity and uniqueness
of this result, since this will be helpful when we construct a scaling model
below. Given that we are studying holomorphic gradient flows automorphic under
$\Gamma_{0}(2)$, up to additive and multiplicative constants this is probably
the only form the potential can take. Holomorphic scalar forms are scarce, and
in fact no such function exists for the full modular group. The obvious way to
build one is by taking a weightless ratio of cusp forms, but for
\textrm{SL}(2,{\textbf{Z}}) ~all cusp forms are built from $\Delta(\sigma)$,
so that there is at most one linearly independent cusp form at each weight,
and no non-trivial scalar can be constructed. Sub-modular flows are a
different matter. In particular, for $\Gamma_{0}(2)$ there are two linearly
independent weight $12$ cusp forms: the oldform $\Delta(\sigma)$ inherited
from \textrm{SL}(2,{\textbf{Z}}) ~and the newform $\Delta(2\sigma)$, and
consequently $\ln\left(  \Delta(2\sigma)/\Delta(\sigma\right)  )$ is a viable
potential\footnote{We believe that any scalar form is generated by this one.}.

This is as far as the \textquotedblleft phenomenological\textquotedblright%
\ approach advocated in \cite{LR} can take us. What remains is to determine
the critical exponents and for this it is necessary to identify an EFT in the
universality class of the quantum Hall system. It is a general result from
conformal perturbation theory that all scaling behaviour is determined by
conformal data, i.e. by the scale invariant (conformal) EFT which exists at
the critical point. It is therefore sufficient to work with scale-invariant
theories to extract the data we need: the RG-potential $C$ and the
Zamolodchikov metric $G$. In general no way is known of extracting the
C-function and metric directly from scaling data, but in our case the
constraints (the symmetry and analyticity in the scaling region) are so strong
that everything is determined up to constants: once a candidate EFT has been
identified our task is reduced to determining normalizations.

The identification of an appropriate effective scaling theory is helped
enormously by the need to have a theory invariant under the sub-modular group
$\Gamma_{0}(2)$. Our experience with conformal field theories immediately
suggests that we consider free fermions on a torus with complex structure
$\tau$ \cite{LAG}. If we do not endow the manifold with any additional
structure, all choices of the lattice defining the torus which are related by
fractional linear transformations in \textrm{SL}(2,{\textbf{Z}}) ~are
equivalent, in the sense that these tori have the same complex structure. The
additional structure we need is a choice of spin structure, of which there are
four on the torus. These are the four possible combinations of boundary
conditions that can be imposed on a fermion along the two cycles of the torus:
PP, PA, AP, and AA, where P denotes periodic and A denotes anti-periodic.

The physical interpretation of these spin structures may be seen from a
standard gauge argument \cite{LH} which forces the Hall conductivities to have
fractional odd denominator values at the IR fixed points. The group which
respects this parity structure is $\Gamma_{0}(2)$, since it groups the
fractions into two equivalence classes: those with odd denominators and those
with even denominators. In other words, the standard (fully spin-polarized)
Hall system contains only one type of anyons, and by exploiting the symmetry
we have grouped the quantum Hall phases, and their anyonic excitations, into a
single equivalence class. Both the PP and PA spin structures are invariant
under $\Gamma_{0}(2)$, but PP is also invariant under the full modular group,
so only PA respects the statistics of the quantum Hall anyons, and is
therefore the extra data we need for our EFT. Since the space of complex
structures (space of inequivalent $\tau$) of this torus model coincides with
the space of conductivities ($\sigma$) observed in experiments, it is clear
that we should identify these spaces and set $\tau=\sigma$.

Finally, both the disorder and the observed antiholomorphic scaling in the QHS
strongly suggests that the scaling should derive from a model with broken
supersymmetry. This has the immediate benefit that we can avail ourselves of
results from superstring theory, where it was shown [10] that Zamolodchikov's
physical metric on the moduli space of complex structures that appears in the
low-energy geometric limit of supersymmetric sigma-models with Calabi-Yau
targets, coincides with the mathematical Weil-Petersson metric. The simplest
Calabi-Yau space is the torus, and in this case the Weil-Petersson metric
reduces to the Poincare metric on the upper half plane. Since this is the
hyperbolic metric of constant negative curvature we have $G^{\sigma
\overline{\sigma}}=(\operatorname{Im}\sigma)^{2}\delta^{\sigma\overline
{\sigma}}.$ We cannot expect this metric to be correct away from the critical
region, but it is sufficient for the purpose of computing the critical exponents.

In summary, we have arrived at a simple EFT which plausibly encodes the
correct anyonics and concommittant symmetries for the quantum Hall system. As
is often the case for effective theories, the problem of deriving the EFT from
microscopics has been sidestepped by exploiting the low-energy symmetries. We
can now calculate the critical exponents in these models, and if they agree
with experiments we will have strong evidence that it is in the universality
class of the quantum Hall system.

We extract the scaling part of the C-function by computing the vacuum energy
$F$ of the EFT, because all scalar potentials, including the real potentials
$F$ and $C$, must be functions of the unique potential $\varphi$ invariant
under the symmetry $\Gamma_{0}(2)$, and hence $C=C(F).$

At criticality both the free energy F and the central charge $c=C(\sigma
_{\otimes},\overline{\sigma}_{\otimes})$ count degrees of freedom. \ Since
this must be true for an arbitrary number of degrees of freedom this implies
that\footnote{This can also be inferred from a standard finite size scaling
argument combined with the constraints of automorphy.} $C(F)\varpropto F$, so
that near the critical point we have $C\approx cF/F_{\otimes}.$

The vacuum energy $F_{PA}=-\ln\mathcal{Z}_{\mathrm{PA}}=-\ln\mathrm{Det_{PA}}$
of PA-twisted fermions on a torus is well known. The determinant factorizes
into holomorphic and anti-holomorphic pieces \cite{LAG}, so we have
$F_{PA}\varpropto f_{PA}+\bar{f}_{PA}$ \ with $f_{PA}(\sigma)=-\ln
\mathrm{Pf_{PA}}(\sigma)$, where the Pfaffian $\mathrm{Pf_{PA}}(\sigma)$ is
the functional \textquotedblleft determinant\textquotedblright\ for a single
PA-twisted spinor on the torus\footnote{Note that the determinant for the
PP-torus vanishes due to the existence of a zero mode, which is consistent
with our previous observation that there is no scalar potential for the full
modular group due to the paucity of cusp forms.}. It is a \textquotedblleft
polynomial\textquotedblright\ in $q=\exp(2\pi i\sigma)$ of infinite degree
which can be extracted from the literature on conformal field theories
\cite{LAG}:
\begin{equation}
\mathrm{Pf_{PA}}(\sigma)=q^{\frac{1}{12}}\prod_{n=1}^{\infty}(1+q^{n})^{2}.
\end{equation}
Recalling the definition of the Dirichlet eta-function:
\begin{equation}
\eta(\sigma)=q^{\frac{1}{24}}\prod_{n=1}^{\infty}(1-q^{n})
\end{equation}
we find that $f_{PA}(\sigma)=\ln\eta(2\sigma)-\ln\eta(\sigma)$. Furthermore,
since $\Delta(\sigma)=\eta(\sigma)^{24}$ we have $f_{PA}=\varphi/24$ and
$F_{PA}\varpropto\operatorname{Re}\varphi$, which gives
\begin{equation}
C(\sigma,\overline{\sigma})\approx c\text{ }\operatorname{Re}\varphi
(\sigma)/\operatorname{Re}\varphi(\sigma_{\otimes}).
\end{equation}

Putting all this together and expanding the beta-function around the critical
(saddle) point at $\sigma_{\otimes}=(1+i)/2$, we find:
\begin{equation}
\beta^{phys}\approx\frac{1}{12}(\operatorname{Im}\sigma_{\otimes}%
)^{2}\overline{\partial}C(\sigma,\overline{\sigma})=\frac{1}{\nu}(\bar{\sigma
}-\bar{\sigma}_{\otimes})+\dots
\end{equation}
with the superuniversal anti-holomorphic scaling exponent $\nu\approx5.2/c.$
\ With two twisted complex fermions ($c=2$) we obtain $\nu=2.6$ which agrees
with scaling data within experimental error \cite{expdata,numdata,LR2}.

In conclusion we have identified a discrete set of EFTs, one of which may be
in the same universality class as the quantum Hall system; i.e. which posesses
the same symmetry properties and scaling behaviour as the low energy transport
theory of the quantum Hall slab. Using the Zamolodchikov form for the
beta-function, which is consistent with the observed anti-holomorphic scaling,
we have calculated the partition function of these effective theories and
extracted the critical exponents. The fact that one of these models, with a
reasonable\ value of c, agrees with the observed values of the exponents
suggests that this effective scaling theory provides an accurate description
of the quantum Hall system near criticality.

\bigskip

\noindent$\mathbf{Acknowledgement}$ \ We are grateful to Philip Candelas and
Andrew Dancer for valuable discussions, and to PPARC for support under grant
no. PP/D000424/1. CAL is indebted to Jesus College, where some of this work
was done.


\begin{thebibliography}{99}                                                                                               %
\bibitem {LR}C.A. L\"{u}tken and G.G. Ross, Phys. Rev. B \textbf{45}, 11837
(1992); \textbf{48}, 2500 (1993).

\bibitem {K}D.E. Kmel'nitskii, JETP Lett. \textbf{38}, 552 (1983).

\bibitem {LLP}H. Levine, S.B. Libby and A.M.M. Pruisken, Phys. Rev. Lett.
\textbf{51}, 1915 (1983).

\bibitem {BL}C.P. Burgess and C.A. L\"utken, Nucl. Phys. \textbf{B500}, 367 (1997).

\bibitem {LR2}C.A. L\"utken and G.G. Ross, Phys. Lett. \textbf{A356}, 382 (2006).

\bibitem {Zam}A. Zamolodchikov, JETP Lett. \textbf{43}, 730 (1986).

\bibitem {LH}R. Laughlin, Phys. Rev. B \textbf{23}, 5632 (1981); B. I.
Halperin, Phys. Rev. B \textbf{25}, 2185 (1982).

\bibitem {LAG}L. Alavarez-Gaum\'e, G. Moore and C. Vafa, Commun. Math. Phys.
\textbf{106}, 1 (1986).

\bibitem {PC}P. Candelas, T. H\"{u}bsch and R. Schimmrigk, Nucl. Phys.
\textbf{B329}, 583 (1990).

\bibitem {expdata}S. Koch, R.J. Haug, K.v. Klitzing and K. Ploog, Phys. Rev.
Lett. \textbf{67}, 883 (1991); H.P. Wei, L.W. Engel and D.C. Tsui, Phys. Rev.
B \textbf{50}, 14609 (1994).

\bibitem {numdata}B. Huckestein, Europhys. Lett. \textbf{20}, 451 (1992);
Phys. Rev. Lett. \textbf{72}, 1080 (1994); J.T. Chalker and J.P.G. Eastmond,
as reported in B. Huckestein, Rev. Mod. Phys. \textbf{67}, 357 (1995).
\end{thebibliography}
\end{document}